# Advances in Micro-Pattern Gaseous Detectors and their Applications

V. Peskov[1,2]
[1]UNAM, Mexico
[2]CERN Geneva, Switzerland
e-mail:Vladimir.peskov@cern.ch

**Abstract**

In the last decade, several new imaging gaseous detectors, called "micro-pattern gaseous detectors", have been conceived and rapidly developed. In this paper, the most popular designs of these detectors and their applications to high energy physics and astrophysics experiments will be described. The main focus of this review will however be on the ongoing activities on practical applications of micro-pattern detectors to medicine and industry

## 1. Micro-Pattern Gaseous Detectors

At the end of the last century, the main imaging gaseous detector, widely used in various scientific researches and applications, was the multi-wired proportional chamber (MWPC) invented in 1968 by G. Charpak [1]. This detector was able to obtain fast electronic images of photons and tracks of elementary charged particles with a 1-D position resolution better than 100 µm. For this great invention, which really revolutionized the detection technique, G. Charpak was awarded in 1992 with the Nobel Prize in physics. However, the triumph of the MWPCs did not last very long. In the beginning of 1990, a new breakthrough happened in the technology of gaseous detectors as a result of the efforts of several teams, the so-called micro-pattern gaseous detectors were developed.

These novel detectors are characterized by the following two features: 1) the gap between the anode and the cathode electrodes is usually very small, sometimes smaller than 50 µm, 2) the electrode structures are manufactured via microelectronic technology enabling to achieve a very high granularity and thus a position resolution much better than in the case of the MWPCs.

Nowadays, there exist quite a lot of various designs of micro-pattern detectors; which can roughly classified into four categories: strip-type, dot-type, hole-type and a parallel-plate type. A detailed description of these detectors can be found in several review articles, see for example [2, 3]; we would like to focus in this paper only on the most popular designs of today: the hole-and the parallel-plate–types.

In a few words, the hole-type detector is a dielectric sheet metalized on both sides in which an array of holes are etched (Figure 1). If a voltage drop V is applied across the metalized surfaces, the field lines will exhibit a focusing effect inside the holes as shown in Figure 2, whereas a very high electric field can be created in this region so that at some critical value of V (V>Vc) the avalanche multiplication of primary electrons happens. This effect was first demonstrated in [4]; later, glass capillary plates were developed [5] and finally F. Sauli suggested a very elegant version of the hole-type detector in which the dielectric sheet was manufactured from a 50 µm thick Kapton [6]. He named this detector: Gas Electron Multiplier (GEM).This latter modification allowed the production of large-area detectors of various shapes. GEM detectors offer very good position resolutions: about 50 µm for photons and even better for tracks of charged particles.

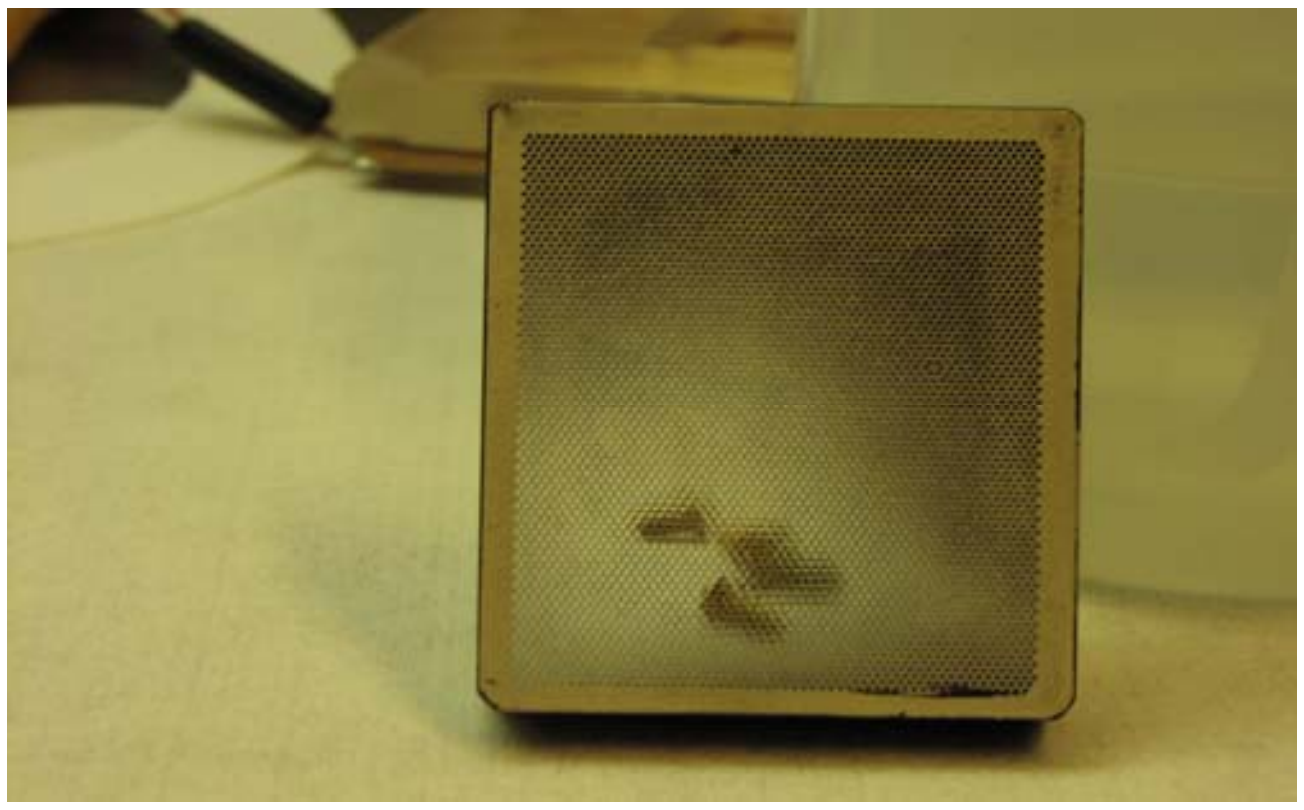

**Figure 1.** A photograph of the first hole-type gaseous multiplier as developed by Del-Guerra et al [1].

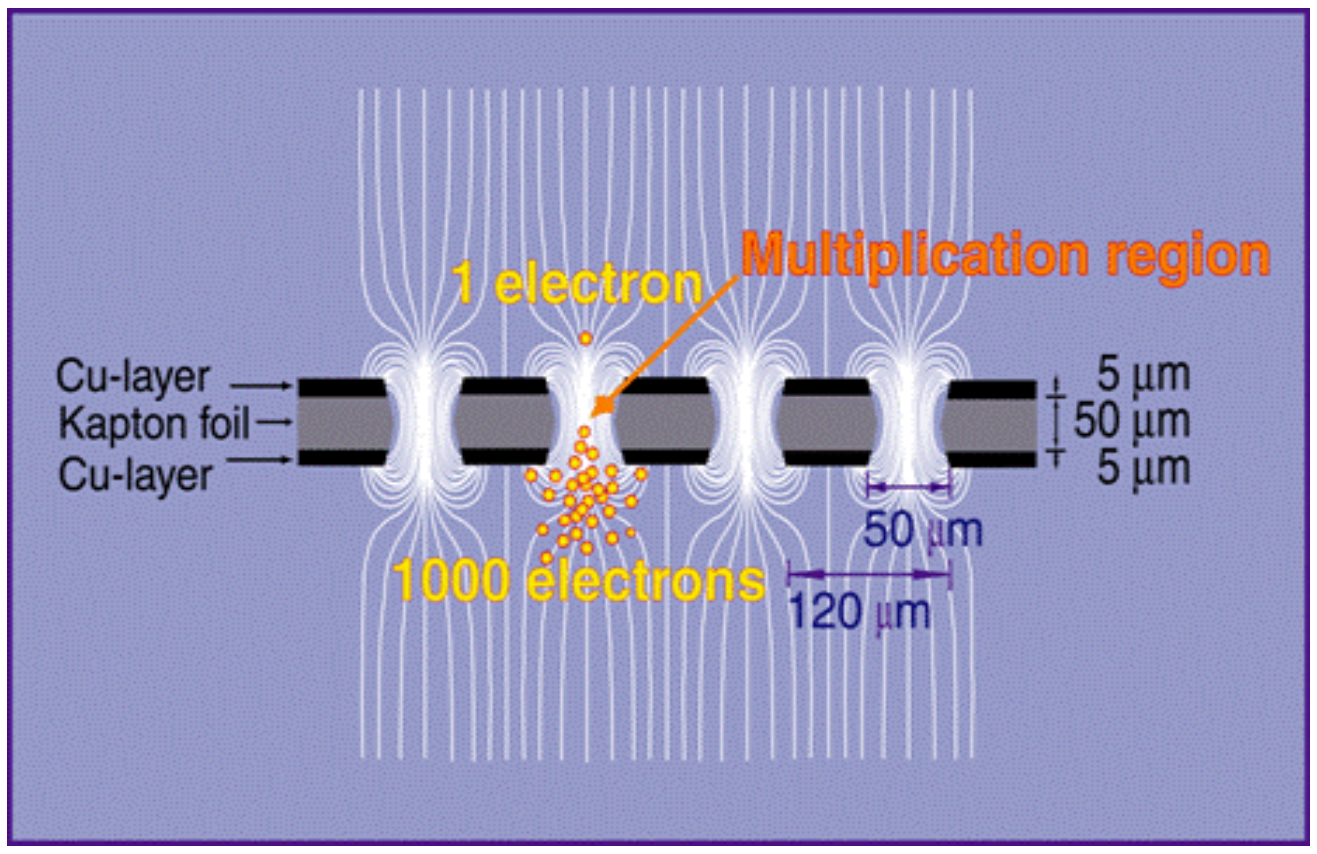

**Figure 2.** The principle of operation of the hole-type gaseous detectors: once a voltage drop is applied between the electrodes, a strong electric field is created inside the holes. If a primary electron enters the hole and the voltage V is above a critical value $V_C$, a Townsend avalanche develops.

For applications requiring modest position resolutions (from sub-mm to few-mm) another version of the GEM detector has been recently developed [7, 8] -a thick GEM (TGEM)-see Figure 3. This robust multiplier is economically manufactured by the standard printed circuit board technique: a CNC machine drills the pattern of sub-mm diameter holes in a two-sided Cu –clad, typically made of a sub-mm thick insulator plate. In a later version of this detector, both electrodes have been made of metallic strips (the strips on one side are perpendicular to the strips on the opposite side) allowing the direct position–sensitive readout of signals created by the avalanches developing in the holes - see Figure 4 [9].

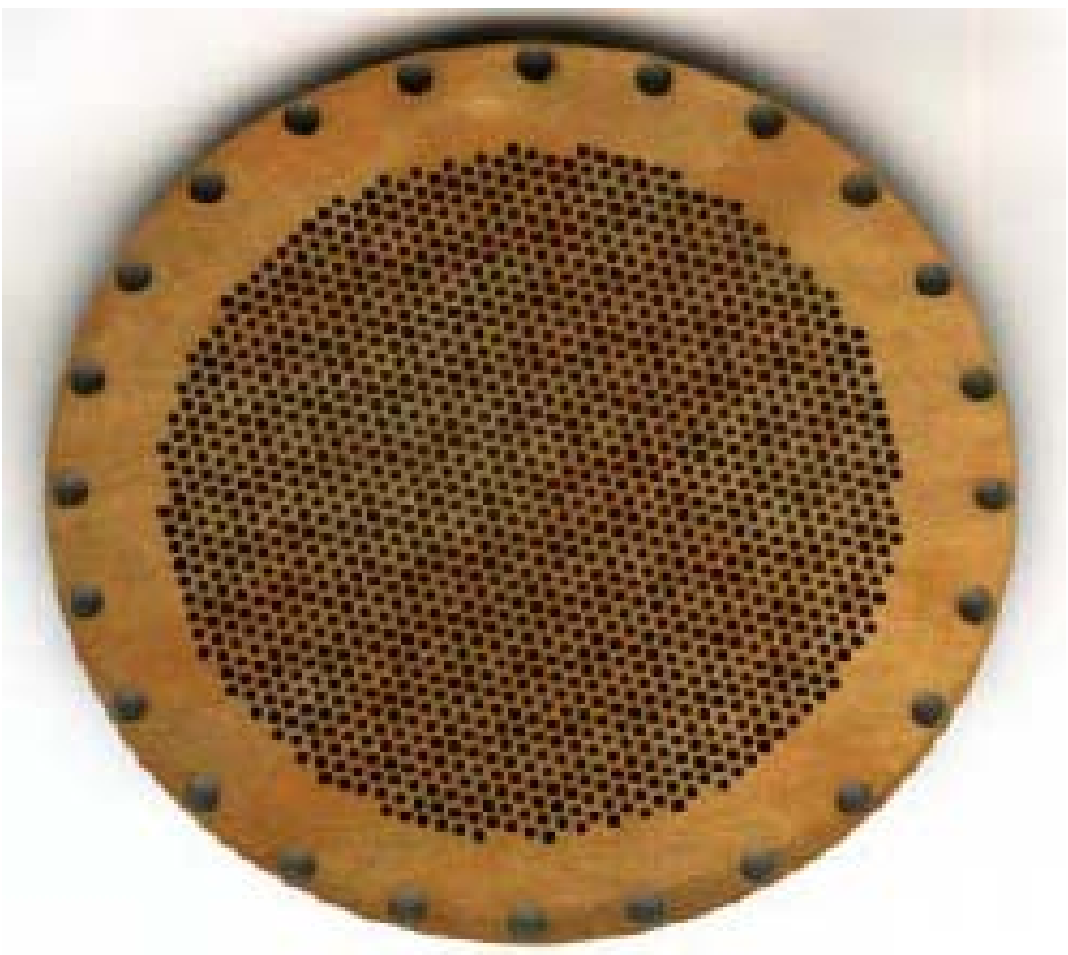

**Figure 3.** Photo of "optimized "GEM [7] named latter as Thick GEM (TGEM)[8].

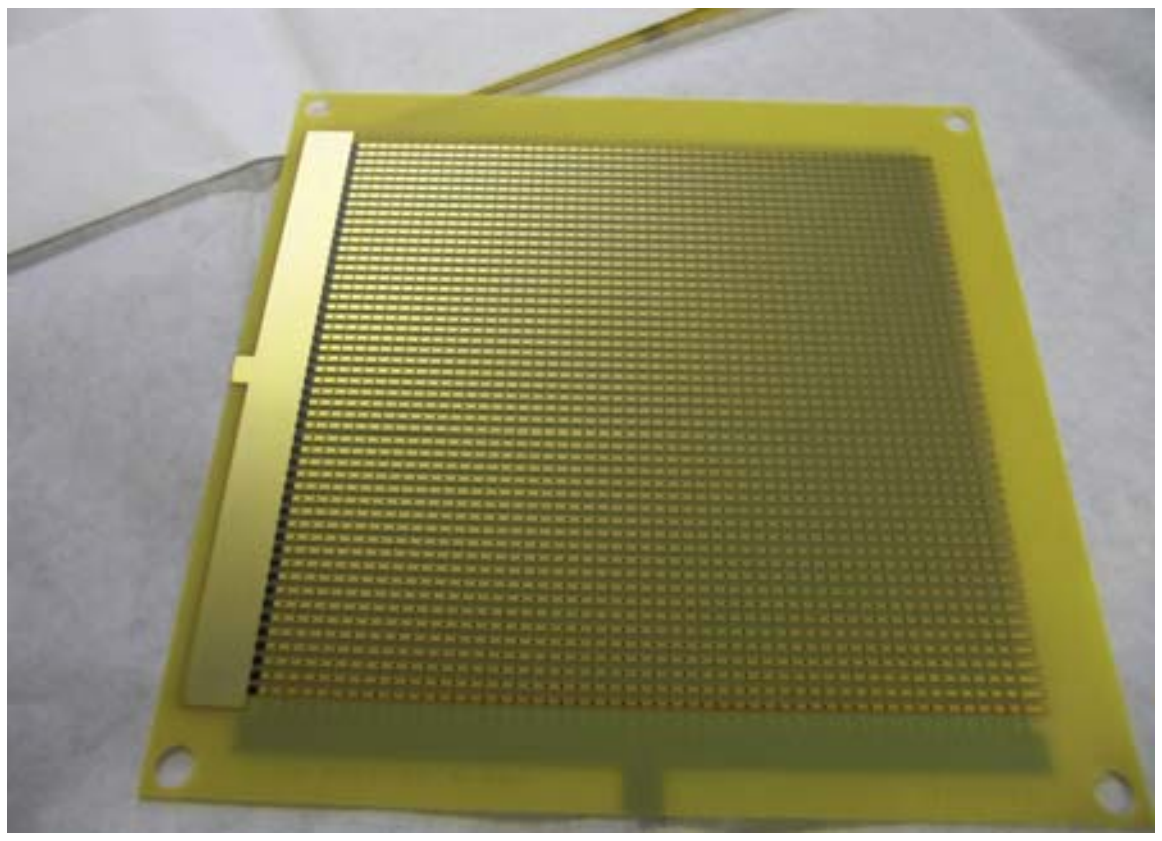

**Figure 4** Picture of a TGEM equipped with a strip electrode.

The parallel-plate-type micro-pattern detectors in principle have the same geometry as the old fashioned parallel-plate gas chambers used for avalanche multiplications at the beginning of the last century (see for example [10]). However, one must notice the following two important modifications: 1) the gap between the electrodes is drastically reduced to 30—300 μm, 2) the anode electrodes have a pattern structure manufactured by a microelectronic technique.

Nowadays, there exist two main designs of parallel-plate-type micro-pattern detectors: a) with cathodes made of a fine mesh and b) with the cathodes made of a metallic imperforated sheet.

The schematic drawing of the mesh parallel-plate-type detector, first developed by I. Giomataris et al [11], is shown in Figure 5. It is called the MicroMesh Gas chamber (MICROMEGAS). In this design the cathode mesh is supported by tiny dielectric pillars between the mesh and the patterned anode. In the latest designs, the gap between the mesh and the anode is maintainned with a high accuracy by placing pillars almost between each opening in the mesh (as shown in Figure 6), therefore this design somehow resembles the GEMs.

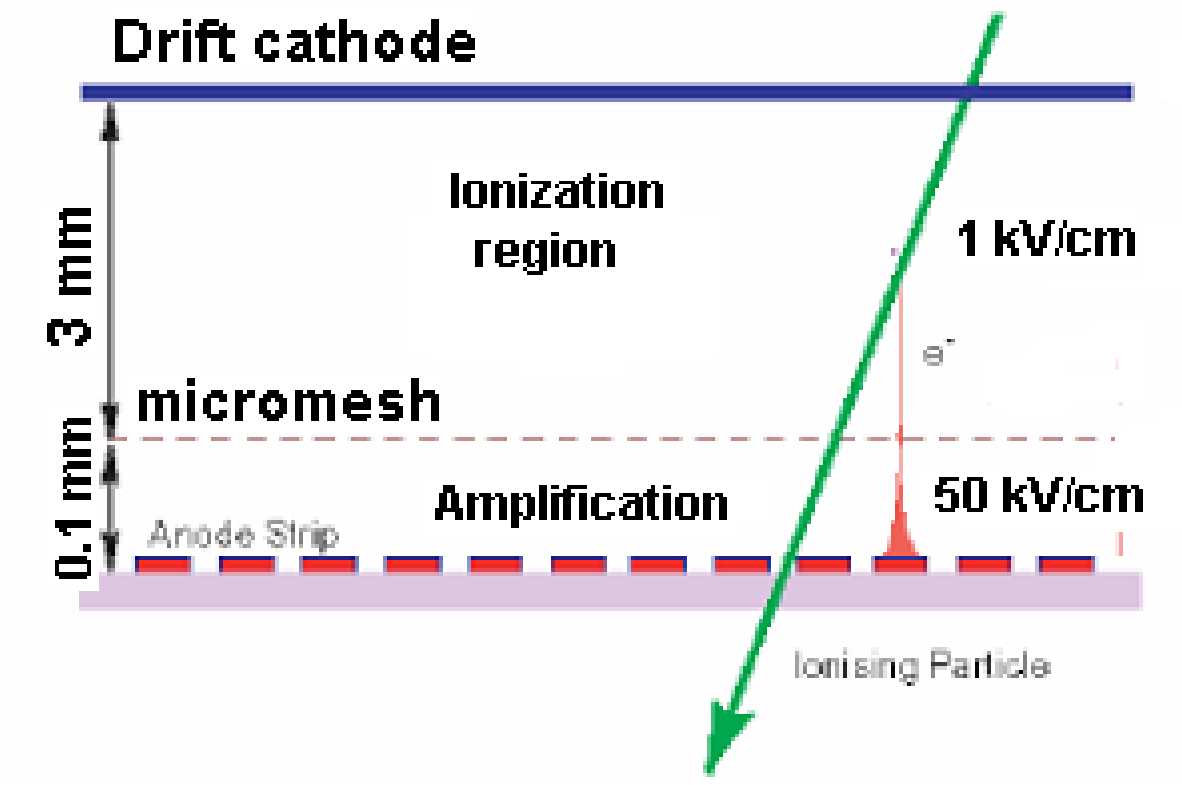

**Figure 5**. A schematic drawing of the MICROMEGAS.

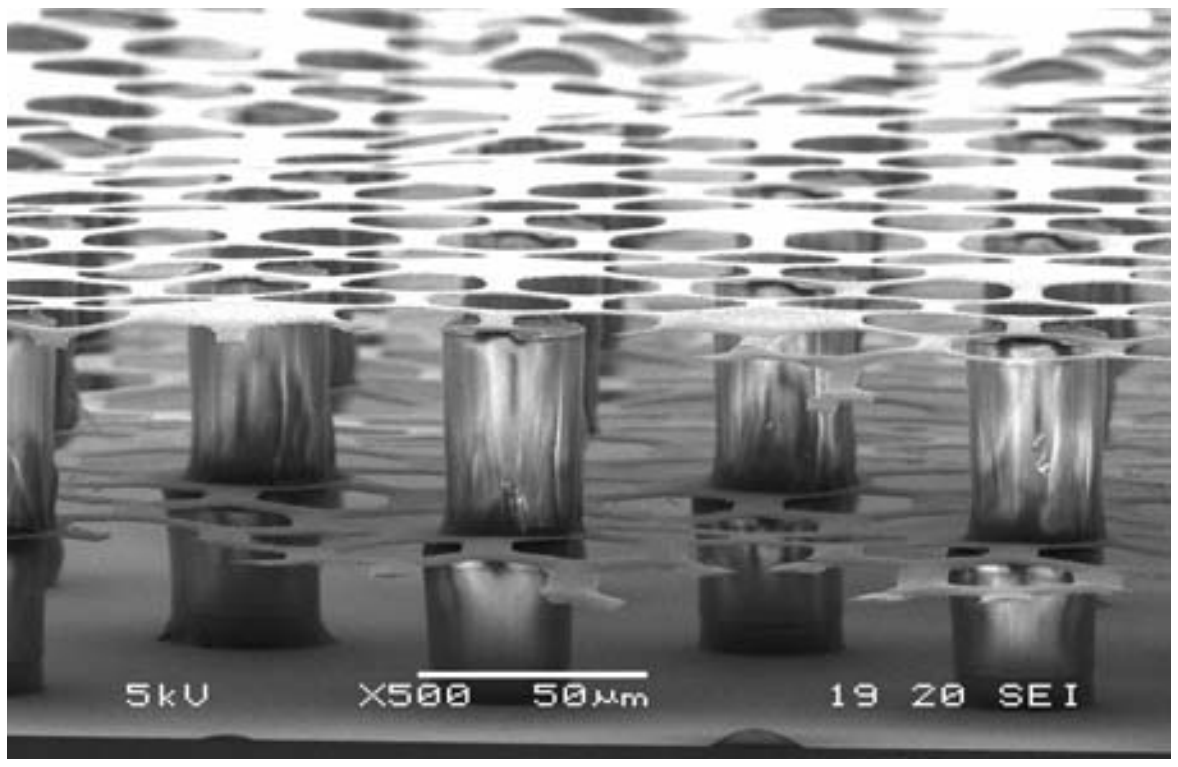

**Figure 6.** Picture of a mesh equipped with supporting pillars as used in modern MICROMEGAS designs.

In advanced designs of GEMs and MICROMEGAS, the patterned anode is directly integrated with the front-end microelectronics making the detector, in the recording of an event, a natural local preamplifier for the electronic circuits. Some electronic systems, for example a multipixel SMOS array or MediPix (see [12, 13] and references therein) allows obtaining time resolved images with a very high position resolution (see Figure 7 and (see [14] and references therein).

The artistic drawing of a parallel-plate-type micro-pattern detector with the metallic imperforated cathode developed by T. Francke et al [15] for medical imaging is shown in Figure 8. In this detector, the glass cathode is coated by metallic strips of 50 μm pitches. By the readout of the avalanche induced signals from these strips, one can obtain high resolution 1-D images. The 2-D images are obtained through the scanning mode featuring the detector movement in the direction perpendicular to the electrode planes.

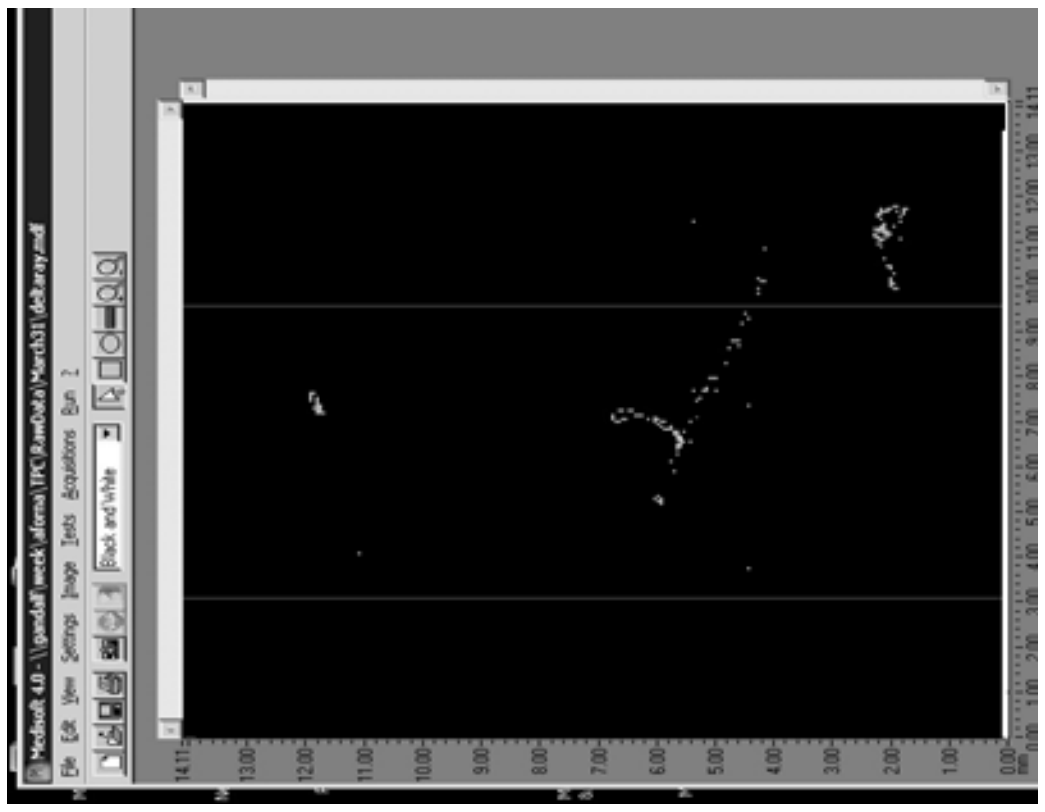

**Figure 7.** Images of delta ray tracks produced by minimum ionizing particles (from [14]).

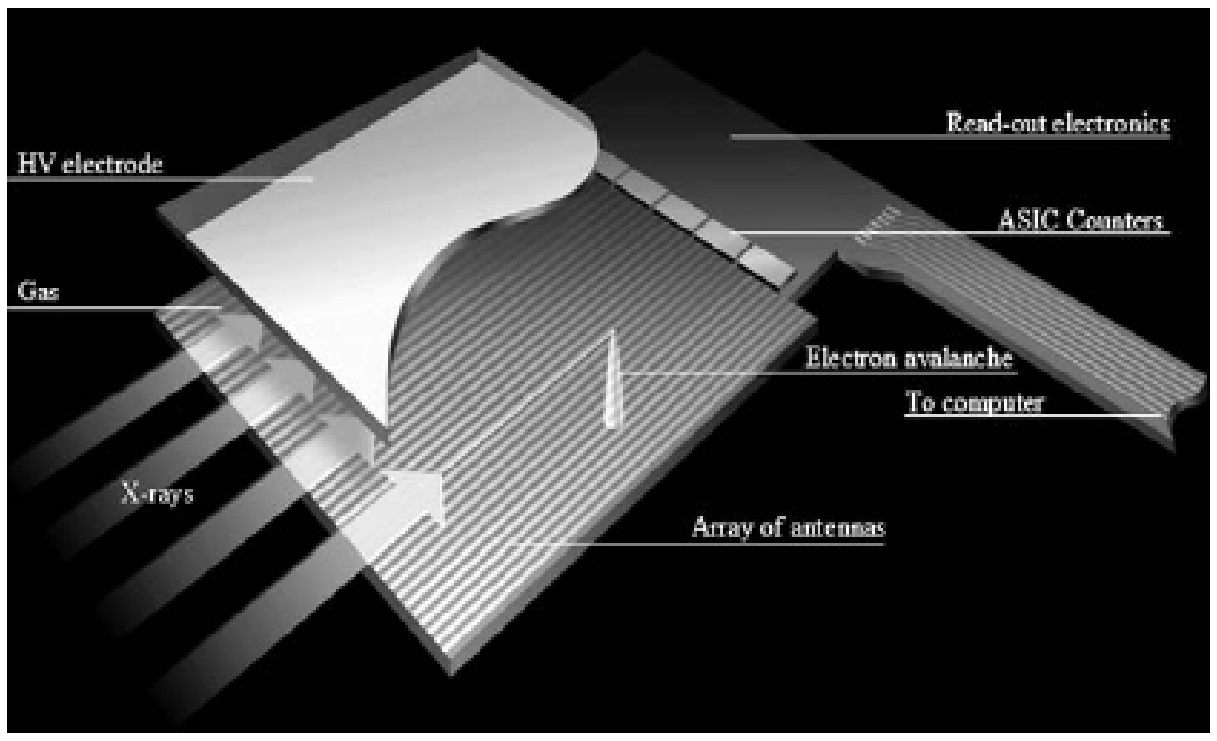

**Figure 8.** An artistic view of a parallel-plate-type micro-pattern detector with the metallic cathode and anode coated with metallic strips.

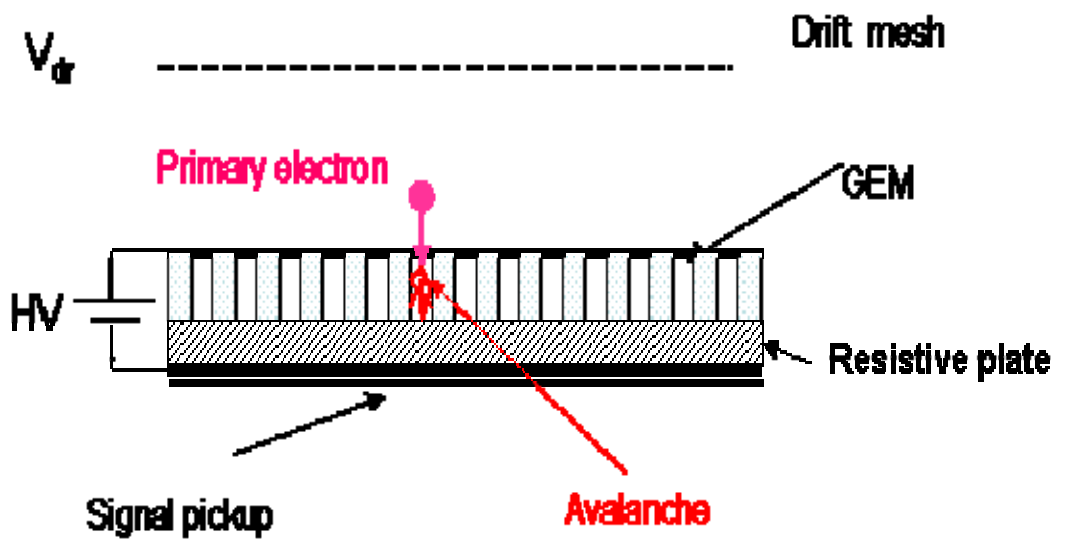

**Figure 9.** One of the early design of GEM with resistive readout plate [20].

Micro-pattern gaseous detectors due to their tiny electrode structure and small avalanche gaps are very fragile and can be easily damaged by sparks appearing at high operational gains (typically at gains of $10^4$ or slightly more). This is why the recent focus of several groups was on the development of more robust designs of micro-pattern detectors. The fist successful attempt was the development of micro-gap resistive chambers. Their designs were similar to the one shown in Figure 8, however, the cathode was made not of metal material but of Pestov glass (resistivity $10^9$-$10^{10}\Omega$cm) or GaAs (resistivity of $10^8$-$10^9$ $\Omega$cm)[16,17]. In the case of sparks, like in usual RPCs, due to the high resistivity of the cathode the spark energy was reduced 100-1000 times and these discharges were not able to destroy the strips or the front-end electronics anymore. This device was successfully used for medical imaging [18].

There is another important feature of this detector: due to small avalanche gaps, the prefect parallel-pate geometry and the capability to operate at gas gains approaching a sparking limit (typically $10^5$-$10^6$) this detector has a potential in achieving an extremely good time resolution. Indeed, with a simplified version of the micro-gap RPC (without fine pitch anode strips) a time resolution of about 50 ps was demonstrated [19].

A resistive electrode approach was successfully applied to other micro-pattern detectors: recently a GEM detector was developed with resistive electrodes [9, 20- 22] (see Figure 9) as well as an InGrid MICROMEGAS with a protective resistive layer [23]. By careful optimization of the thickness and the resistivity of the resistive protective layers one can achieve at the same time a spark-protective effect and a high rate capability [16, 24].

Since today this is the only way to ensure the spark protection, one can foresee that the next generation of micro-pattern detectors will be based on various resistive electrode microstructures.

## 3. Micro-Pattern Photodetectors and their Applications

Nowadays, micro-pattern gaseous detectors are used in many applications ranging from high energy physics and astrophysics experiments to some practical applications.

Reviews of applications of micro-pattern detectors in high energy physics and astrophysics can be found in several papers (see for example [25, 26]). Basically, they are used as high position resolution detectors in various charged particle tracking devices including TPCs and recorders of the micro tracks produced by photoelectrons in some polarimeter designs. It is relevant, however, to stress another very important application of micro-pattern detectors in high time resolution detectors and in photodetectors. For example, the ALICE TOF and HADES TOF detectors consist of arrays of micro-gap RPCs briefly described above. GEMs coated with CsI layers are used in hadron blind detector at RHIC-PHENIX and resistive GEMs are considered as triggering devices and as photosensitive elements for the upgraded ALICE RICH [27].

In the following, however, we would like to focus on the description of very promising applications of micro-pattern detectors in medicine, plasma diagnostics and high resolution UV detectors.

### 3.1 Perspectives in Medical Applications of Micro-pattern Detectors

Owing to their good position and time resolutions, the application of micro-pattern gaseous detectors in medicine is extremely attractive. We will describe below only some examples.

#### 3.1a Portal Imaging

About half of the cancer patients in the world are today treated with radiotherapy. During the treatment, it is of great importance to closely monitor the local dose delivered to the tumor and the surrounding tissue to ensure an effective destruction of the cancer cells while the healthy ones are not damaged. This goal can be accomplished with the help of the so-called electronic portal imaging device. The basic idea of this device is to combine in one flat panel package two imaging devices: one for obtaining a continuous x-ray image of the tumor and the second one to monitor the position of the therapeutic beam-see Figure 10. This allows the on line registration of the beam misalignment and a fast automatic correction of its position. Hole-type micro-pattern detectors combined with solid gamma converter offer a simple and cost effective solution [18, 28]. One of such prototypes, developed and successfully tested by a Swedish team, is shown in Figure 11. It is a sandwich of perforated solid convertors and GEM detectors. In the case of the keV diagnostic imaging, photons absorbed in the gas create primary electrons which are then multiplied by the GEM and drift through the holes of the solid converters to the pixel readout plate. In the case of the MeV therapeutic imaging, Compton electrons and electron-positron pairs from the solid converter produce the ionization in the gas volume whereas the resulting electrons also drift to the readout board. This enables to obtain simultaneously x-rays and gamma images. As an example, Figure 12 shows the bone surrounded by a tissue recorder by the GEM-based prototype of the electronic portal imaging device. The commercial prototype of this detector is currently under development by several companies. A similar detectors concept is also developing by Sauli group [29].

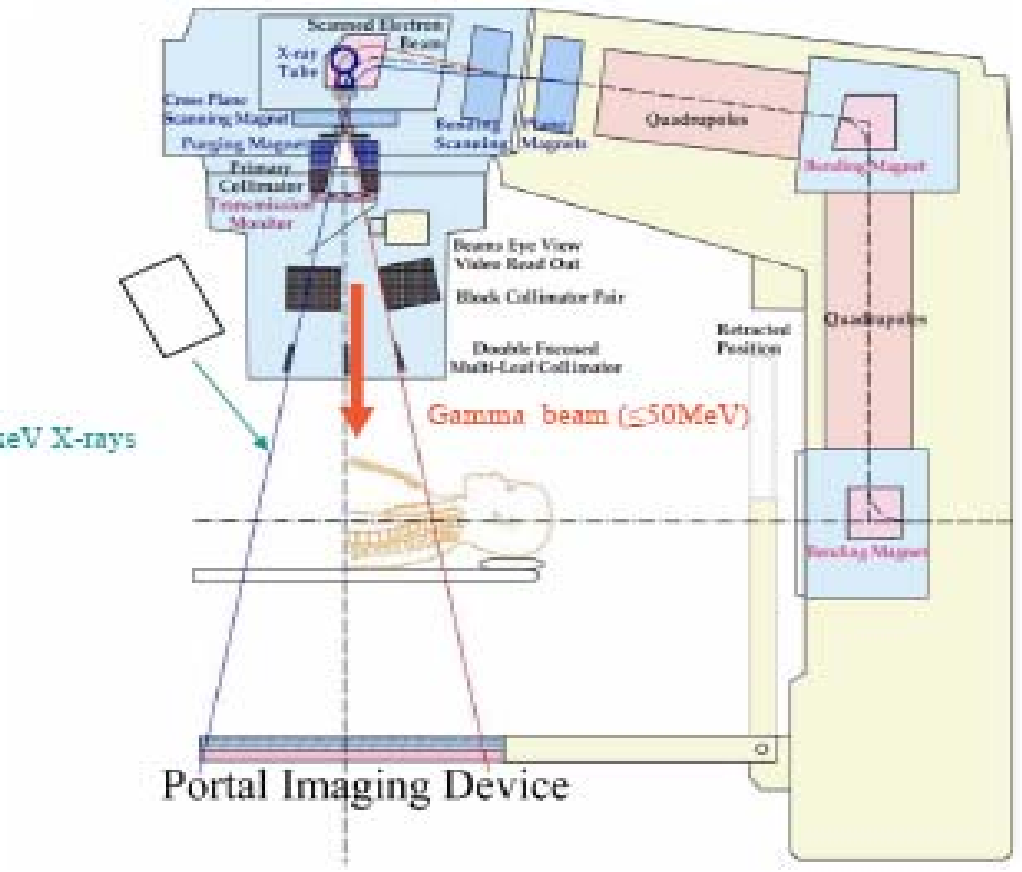

**Figure 10.** A schematic drawing of the cancer treatment machine for gamma therapy of patients at the Karolinska hospital in Stockholm. One can see an accelerator producing a gamma beam, a patient and a portal imaging device.

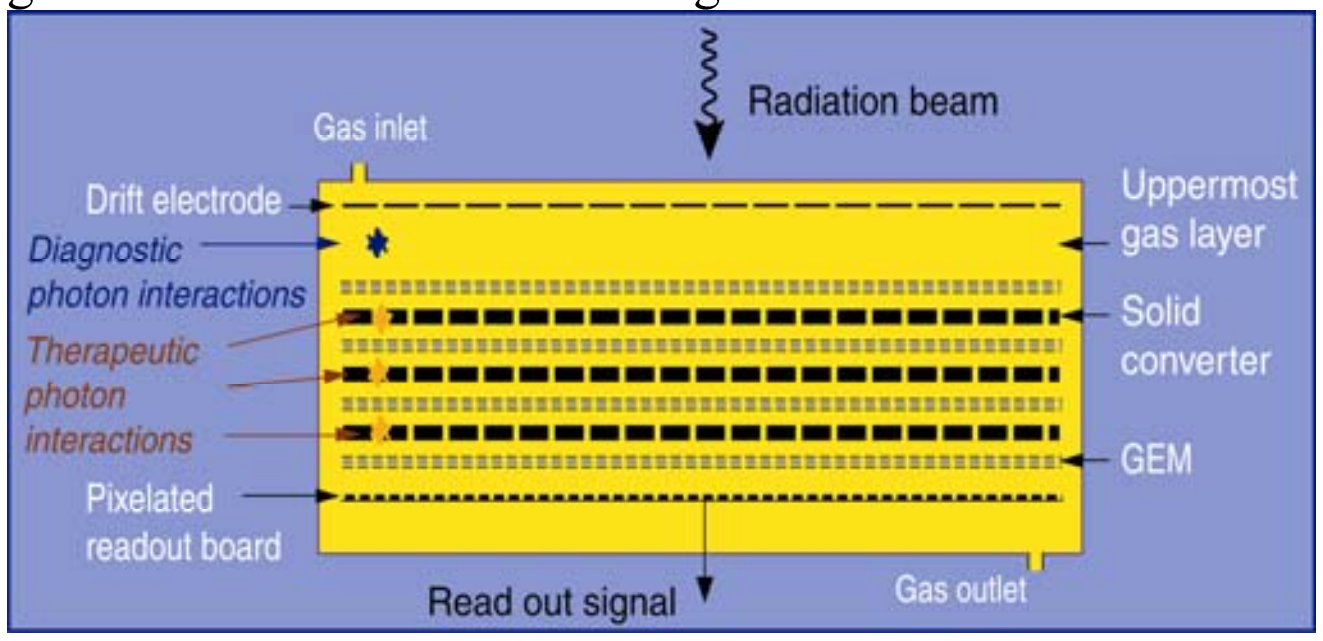

**Figure 11.** A schematic drawing of the GEM-based electronic portal imaging device. It is a sandwich of GEMs and perforated metallic convertors combined with a pixelized readout plate.

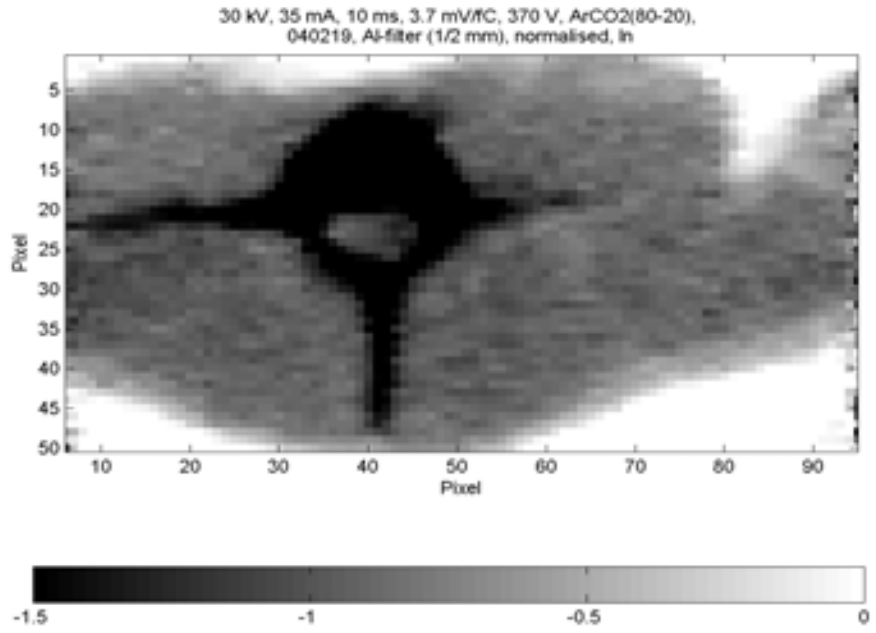

**Figure 12**. An image of the bone and the surrounding tissue as obtained by the GEM-based portal imaging detector shown in Figure 10.

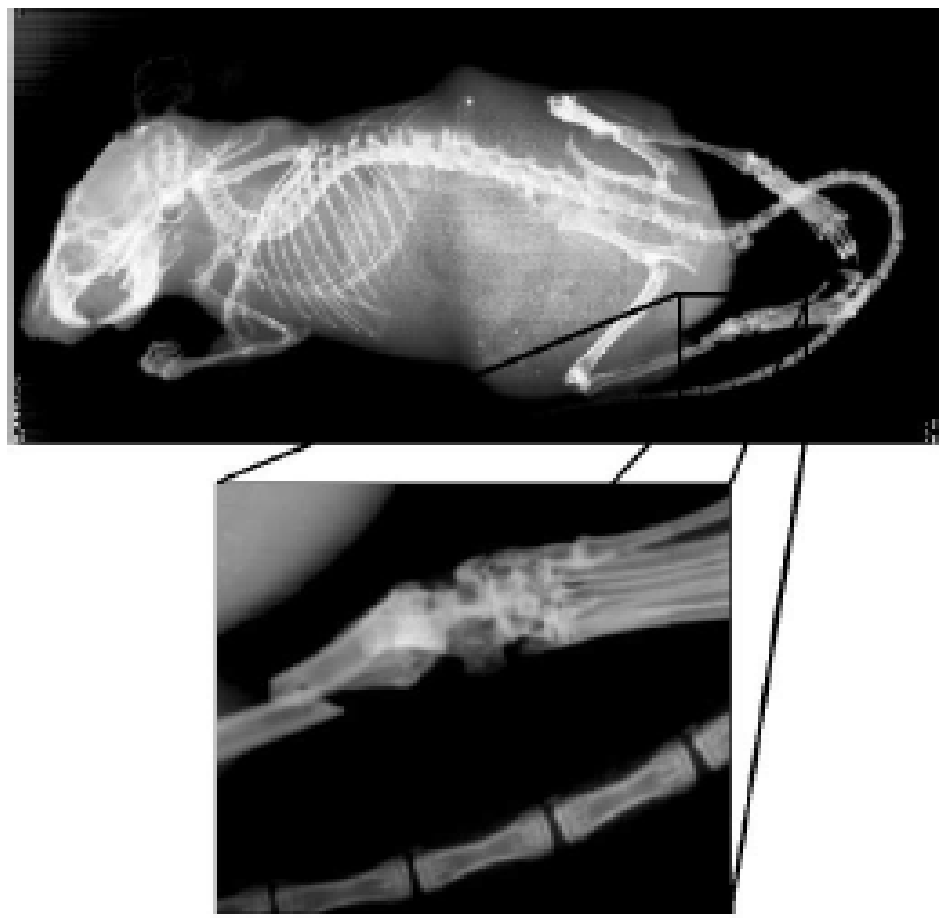

**Figure 13.** An image of a small animal as obtained with a parallel-plate-type micro-pattern detector.

### 3.1b. X-ray Scanners

Parallel-type micro-pattern detectors with metallic and resistive cathodes and strip anodes were successfully used for medical imaging including mammography [15, 18, and 30]. The main advantage of these detectors operating in photon counting/digital modes is a 10 times lower dose delivered to the patient for obtaining the same quality of images than with the standard mammographic machines. As an example in Figure 13 one of the images is shown obtained with such a scanner.

### 3.1c. TOF PET

Conventional PET scanners lack time of flight capabilities and thus project full-length lines of responses for each annihilation event recorded. This introduces a significant amount of noise to the resulting image. Recent developments of parallel-plate-type micro-pattern detectors offer a cheap possibility in building a TOF-PET. Although the TOF-PET allows only negligible improvement to the resolution, it has huge potential for noise reductions. With better time resolution coincidence time windows may be trimmed to reduce random coincidence detections. More important, TOF information allows shorter lines of response to be projected for each annihilation event effectively reducing the amount of noise that is inherent to the modality. Simplified prototypes of such TOF-PETs were already built and successfully tested by the Coimbra group [19, 31].

### 3.1d. Plasma Diagnostics

The MWPCs are used for plasma diagnostics in X-rays and UV regions of spectra since a long time ago [32]. Micro-pattern gaseous detectors, due to their better imaging capabilities, allow the achievement of the same task much more effectively. For example, an innovative system which combines very fast 2D imaging capabilities with spectral resolutions in the X-VUV range 0.2–8 keV have been developed at ENEA-Frascati (Italy) in collaboration with INFN-Pisa (Italy). The full system has been tested on the Frascati Tokamak. This was upgraded in 2001 and on the National Spherical Tokamak experiments (NSTX) in 2002 as a powerful diagnostic tool for magnetic fusion plasmas [33].

### 3.1e. Micro-Pattern UV Sensitive Detectors

Micro-pattern gaseous detectors combined with reflective or semi-transparent photocathodes allow for the obtaining of high resolution images in UV or the visible region of spectra.

First prototypes of such detector, sensitive from UV to visible light, were based on capillary gas multipliers operating at an atmospheric pressure [34]-see Figure 14. Operations at one atm allow the building of large sensitive area detectors since there are no mechanical constrains on the window size and this is why this approach immediately attracted great attention. Nowadays, several groups are developing gaseous photomultipliers based on various micro-pattern detectors (see a review paper [35]). As an example, in Figure 15 a photograph of a sealed GEM based gaseous photomultiplier is shown [35]. A high position resolution image can be obtained with gaseous photomultipliers having the strips or pixel types of the readout plate [36, 37] see Figure 16.

There are many potential applications of such position sensitive photomultipliers. We will mention here only one of them - hyper spectroscopy. The hyper spectroscopy is a new method of surface imaging that simultaneously provides both high position and spectral resolution, thus permitting the remote study of the chemical composition of the surfaces (see [38] and references therein).

As an example Figure 17 shows a hyper spectroscopic image of tundra. Recently UV sensitive micro-gap RPCs were used to extend the hyper spectroscopic method into the UV region. The very first promising results are described in [38].

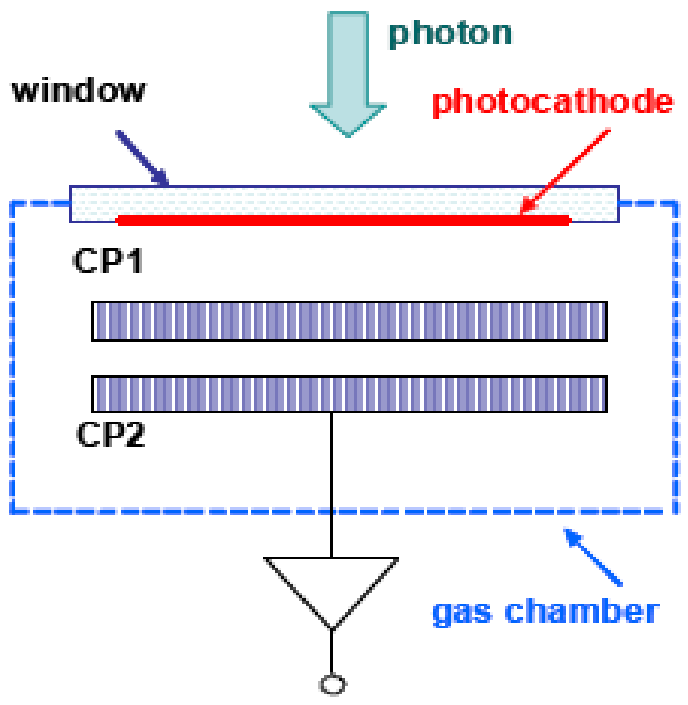

**Figure 14.** A schematic drawing of the capillary-based gaseous photomultiplier sensitive to visible light.

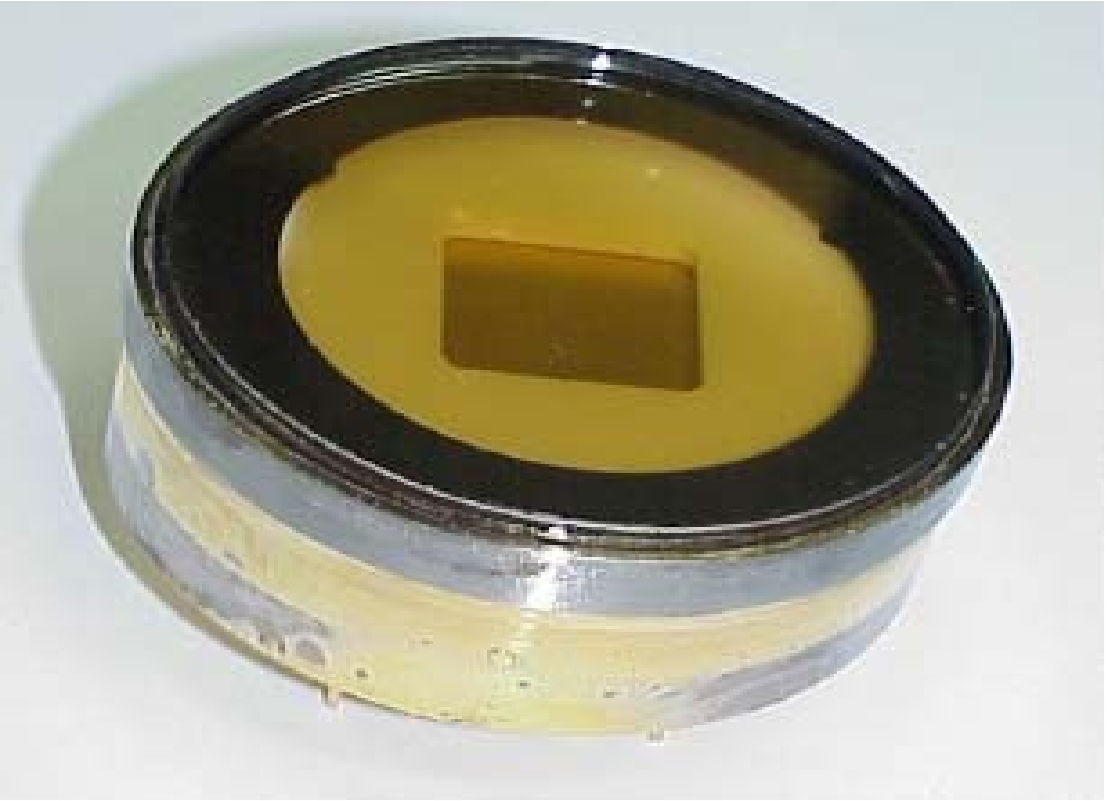

**Figure 15.** A photograph of a sealed GEM- gaseous detector with a semitransparent K-Cs-Sb photocathode.

Another important practical application of photosensitive – micro-pattern detectors is remote flame detection. For example, recently position-sensitive GEMs with resistive electrodes were proposed for early forest fire detection [39].

Finally, it deserves to mention an application of photosensitive hole-type micro-pattern detectors for recording primary electrons and scintillation light from noble liquid detectors. Some fresh results and the brief review of the history of these developments one can find in [9].

## 4. Conclusions

Micro-pattern gaseous detectors are very promising high position-and high time resolution imaging detectors capable to visualize not only tracks of charged particles, but also photons with energies from eV to several MeVs. Due to its unique features, the micro-pattern gaseous detector can find more and more applications.

## Acknowledgments

The author would like to thank Prof. E. Nappi who participated in many activities described here and who has also kindly read and discussed this paper.

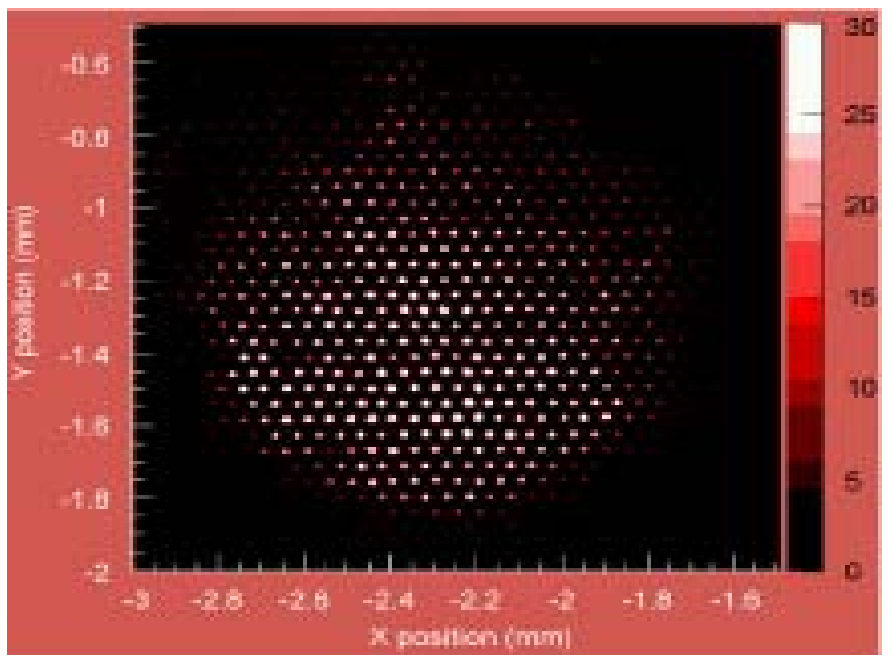

**Figure 16.** An image of GEM holes (50 μm pitch) obtained with GEM-bases UV sensitive imaging detector

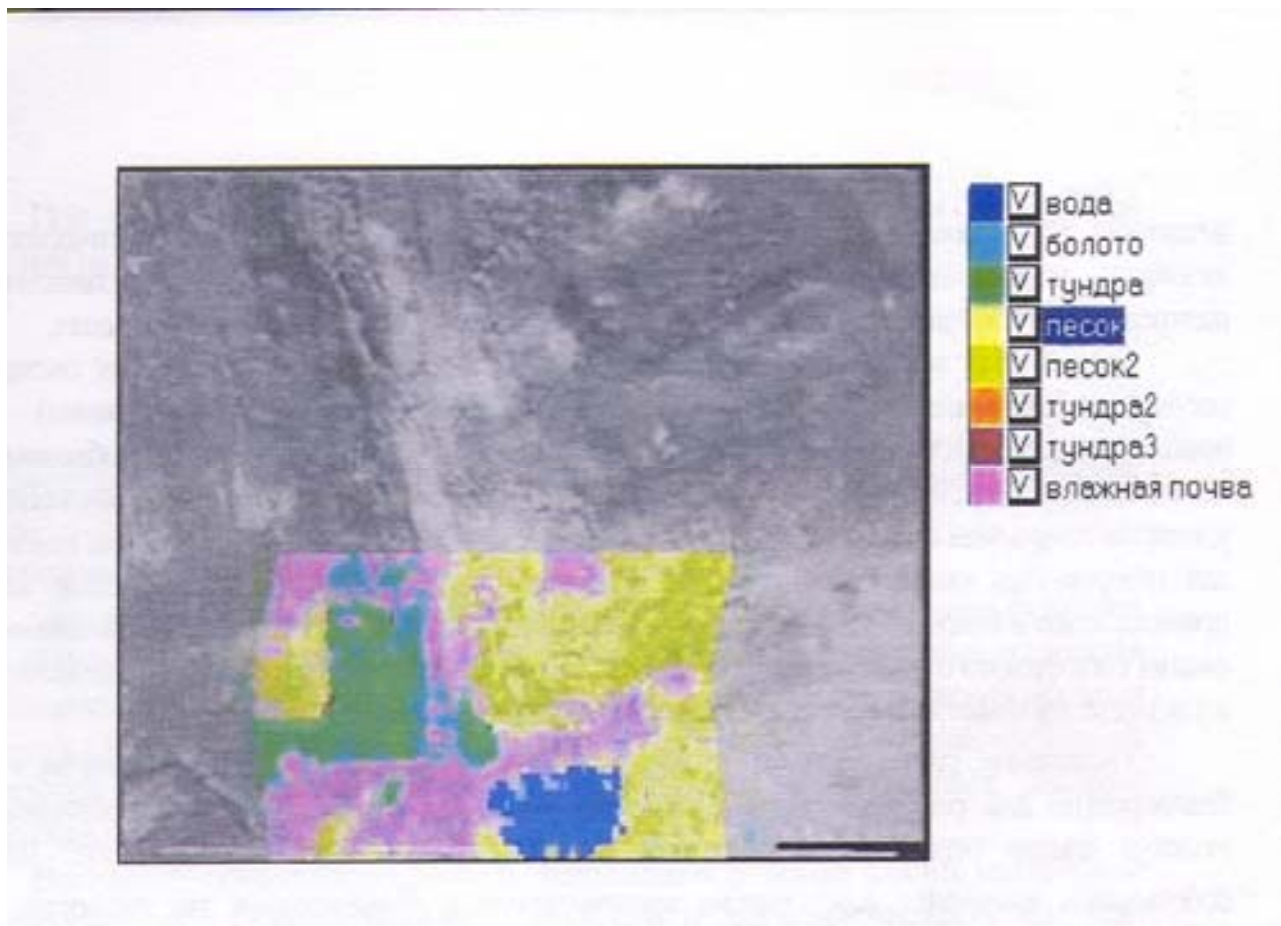

**Figure 17**. A hyper spectroscopic computer synthesized image of the earth's surface. One can identify the surface composition, e.g. the yellow color is sand, the rose color is wheat on earth, blue is water and the light blue represents to swamps.